\documentclass[12pt]{article}
\usepackage{pdproc} 
\usepackage{epsfig}
\begin{document}   

\renewcommand{\thefootnote}{\alph{footnote}}

\newcommand{\ba}{\begin{eqnarray}}
\newcommand{\ea}{\end{eqnarray}}
  
\title{A POSSIBLE QUANTUM-GRAVITATIONAL ORIGIN OF THE NEUTRINO 
MASS DIFFERENCE ? \\
Consequences and Experimental Constraints }

\author{NICK E. MAVROMATOS\footnote{Conference Speaker}}

\address{Department of Physics, King's College London, 
University of London, \\ Strand, 
London, WC2R 2LS, United Kingdom \\
 {\rm E-mail: nikolaos.mavromatos@kcl.ac.uk  }}

  \centerline{\footnotesize and}

\author{SARBEN SARKAR}

\address{Department of Physics, King's College London, 
University of London, \\ Strand, 
London, WC2R 2LS, United Kingdom \\
 {\rm E-mail: sarben.sarkar@kcl.ac.uk  }}

\abstract{We discuss the theoretical 
possibility that the neutrino mass differences
have part of their origin in the quantum-decoherence-inducing
medium of space-time foam, which characterises some models of 
quantum gravity, in much the same way as the celebrated MSW effect,
responsible for the contribution to mass differences when neutrinos 
pass through ordinary material media. We briefly describe 
consequences of such decoherent media in inducing CPT 
violation at a fundamental level, 
which would affect the neutrino oscillation probability; 
we speculate on 
the connection of such phenomena with the r\^ole 
of neutrinos for providing one possible source of a cosmological constant 
in the Universe, of the phenomenologically right order of magnitude. 
Finally we discuss possible experimental 
constraints on the amount of neutrino mass differences induced
by quantum gravity, which are 
based on fits of a simple decoherence 
model with all the currently available neutrino data.}
   
\normalsize\baselineskip=15pt

\section{Introduction and Motivation}

Recent astrophysical observations, using different experiments
and diverse techniques, seem to indicate that 70\% of the 
Universe energy budget is occupied by ``vacuum'' energy density of 
unknown origin, termed Dark Energy~\cite{snIa,wmap}. 
Best fit models give the positive cosmological {\it constant} Einstein-Friedman
Universe as a good candidate to explain these observations, although
models with a vacuum energy relaxing to zero (quintessential, i.e. 
involving a scalar field which has not yet 
reached the minimum of its potential) are 
compatible with the current data.

{}From a theoretical point of view the two categories of Dark Energy models 
are quite different. If there is a relaxing cosmological vacuum energy,
depending on the details of the relaxation rate, it is possible in general
to define asymptotic states and hence a proper scattering matrix (S-matrix) 
for the theory, which can thus be quantised canonically. 
On the other hand, Universes with a 
cosmological {\it constant} $\Lambda > 0$  (de Sitter) 
admit no asymptotic states, as a result of the Hubble horizon which 
characterises these models, and hampers the definition of proper asymptotic
state vectors, and hence, a proper S-matrix. 
Indeed, de Sitter Universes will expand for ever, and eventually
their constant vacuum energy density component will dominate
over matter in such a way that the Universe will enter again an 
exponential (inflationary) phase of (eternal) accelerated expansion,
with a Hubble horizon of radius $\delta_H \propto 1/\sqrt{\Lambda}$. 
It seems that the recent astrophysical observations~\cite{snIa,wmap} 
seem to indicate
that the current era of the universe is the beginning of such an 
accelerated expansion.

Canonical quantisation of field theories in de Sitter space times is 
still an elusive subject, mostly due to 
the above mentioned problem of defining a proper S-matrix.
One suggestion towards the quantisation of such systems could be 
through analogies 
with open systems in quantum mechanics, interacting with an environment. 
The environment in models with a cosmological constant would consist of field modes
whose wavelength is longer than 
the Hubble horizon radius. 
This splitting was originally suggested by Starobinski~\cite{staro},
in the context of his stochastic inflationary model, and later on was adopted
by several groups~\cite{coarse}. 
Crossing the horizon in either direction
would constitute interactions with the environment. 
An initially pure quantum state in such Universes/open-systems would 
therefore become eventually mixed, as a result of interactions 
with the environmental modes, whose strength will be controlled by the 
size of the Hubble horizon, and hence the cosmological constant. 
Such decoherent evolution could explain the classicality of the early Universe 
phase transitions~\cite{rivers}
(or late in the case of a cosmological constant). The approach is still far from being 
complete, not only due to the technical complications, 
which force the researchers to adopt severe, and often unphysical
approximations, but also due to conceptual issues, most of which
are associated with the back reaction of matter onto space-time,
an issue often ignored in such a context. It is our opinion
that the latter topic plays an important r\^ole in the evolution
of a quantum Universe, especially one 
with a cosmological constant, and is associated
with issues of quantum gravity. The very origin of the cosmological constant,
or in general the dark energy of the vacuum, is certainly a property
of quantum gravity.

This link between quantum decoherence and a cosmological constant 
may have far reaching consequences for the phenomenology of 
elementary particles, especially neutrinos.
In this talk we shall elaborate on a scenario~\cite{bm2}, 
suggested originally in \cite{bm2}
according to which the mass differences of neutrinos may have (part of)
their origin in the quantum gravity decoherence medium of space time foam.
The induced decoherence, then, will affect their oscillation,
a notable consequence being the appearance of 
intrinsic CPT violating damping terms in front of the oscillation amplitudes. 
This fundamental (and local) form of 
CPT violation has its origin in the ill-defined nature of 
the corresponding  CPT operator
in such decoherent quantum theories,
due to a mathematical theorem by 
Wald~\cite{wald}. 
This local form of CPT violation, as a result of the 
interaction of the elementary particle with a 
decoherent medium 
is linked to a cosmological
(global) violation of CPT symmetry of the type proposed in \cite{mlambda}
by means of a generation of a cosmological constant as a result 
of neutrino mixing and non-trivial mass differences due to 
the quantum gravity vacuum. The framework in which such 
a cosmological constant may be generated by the neutrinos is the 
approach of \cite{vitiello}, according to which 
the problem of mixing 
in a quantum field theory is treated by means of a canonical
Fock-space quantization. 

The structure of this talk is the following: 
in the next section we discuss the quantum-gravitational 
generation of neutrino mass differences, in 
analogy with the celebrated
MSW effect~\cite{msw}, associated with enhancement 
of oscillations during the passage of neutrinos
through ordinary matter. 
We pay particular attention to discussing 
the associated decoherence effects that 
would characterise the neutrino oscillation formula
in such cases. 
In section 3 
we review a preliminary 
discussion~\cite{bmsw} on the constraints from data for the proportion of neutrino mass differences that can be attributed to 
quantum gravity. The constraints are
imposed by fitting some simplified decoherence models to 
the currently data on neutrinos. 
The data seem to exclude the possibility that the decoherence induced by  
certain types of stochastic space-time foam~\cite{ms} can be 
the exclusive source for the ``observed'' damping in front of the 
oscillation amplitudes in the respective oscillation probabilities 
that fit the data, and consequently imply that, at most, 
only a small
percentage of the mass difference could be of unconventional 
origin due to the space-time foamy medium. 
In section 4, there are speculations on 
a possible link of this foam effect to the generation 
of a cosmological constant in the Universe, of the phenomenologically right 
order of magnitude.
Conclusions are presented in section 5.

\section{Quantum-Gravitational MSW effect and induced decoherence}

In \cite{bm2} the idea that the observed mass differences
between neutrinos are due to
a type of stochastic space-time foam
has been proposed.  The concept presented is the
possibility of the creation of microscopic charged black/white
hole pairs out of the vacuum which would induce information loss
and from their subsequent Hawking radiation would create a
medium with 
stochastically fluctuating electric charges. The microscopic black holes
would radiate preferentially the lightest charged particles i.e. electron/positron pairs 
and the `evaporating' white
hole could then absorb, say, the positrons.  
The resulting electric current fluctuations 
would interact  
non-trivially with $\nu_e$ and not $\nu_{\mu}$, 
according to coherent scattering interactions of the standard model,
resulting in oscillations, and hence, 
effective mass differences, for the neutrinos,
similar to the celebrated MSW effect~\cite{msw} for neutrinos 
in ordinary media.   
We have emphasized the r\^ole of the charged black holes 
in this effect since, from semi-classical arguments given below 
non-charged black holes may have a shorter lifetime. 
This leads us to consider that the effect of space-time foam on
neutrinos can be treated 
similarly to the celebrated MSW effect~\cite{msw} for neutrinos 
in ordinary media. 

Before proceeding with such a MSW-like parametrisation 
of these stochastic quantum-gravity-induced effect, we consider it as
useful to discuss briefly some properties of 
charged black holes which have been derived semi-classically; we will 
extrapolate such results to the case of microscopic space-time foam. 
Owing to the lack of a complete theory of quantum gravity (QG) such an extrapolation cannot be rigorously justified.

Charged black holes can be divided into two kinds: extremal, for which
there is an equality between the electric charge and the mass of the black 
hole, $Q=M$, and the non extremal ones.
According to studies of scalar particles in the background of charged black 
holes~\cite{gao}, extremal black holes do not radiate particles.
Moreover in string theory one can construct
black hole configurations out of stringy membranes by invoking 
appropriate duality transformations, and so 
obtain many properties of non extremal black holes from 
extremal ones in a smooth way~\cite{lifschytz}. 

Such stringy studies have shown that the rate of change of 
the energy (mass $M$) of the near-extremal black holes, is given by
\begin{equation} 
\frac{dM}{dt} \sim \frac{A}{G_N}T^2 
\label{rateofchange}
\end{equation}
where $A$ is the are of the horizon, $T$ is the temperature 
and $G_N$ the gravitational constant.
The above formula demonstrates, therefore, how (stringy) black holes,
viewed as membranes, thermalise. 
It also shows that an extremal black hole, for which $T = 0$, cannot radiate
particles. This last result is also recovered in field theoretic studies
of black holes~\cite{gao}, by actually considering 
the number $N_{\omega_0}$ of massless (scalar) particles 
(or pairs of particles/antiparticles) created in a state 
represented by 
a wavepacket centered around an energy $\omega_0$, is bounded:
\begin{equation} 
N_{n\omega_o\ell m} \le \frac{2c(\omega_0)|t(\omega_0)|^2}{(2n\pi)^{2k_B -1}}
\label{particlecreation}
\end{equation}
where $c(\omega_0)$ is a positive function, $k > 0$ is an arbitrary but large power,  
$\ell, m$ are orbital angular momentum quantum numbers (arising from spherical harmonics in the wavefunction of the packet), and $2n\pi$, $n$ 
positive integer, is a special  
representation of the retarded time 
in Kruskal coordinates~\cite{gao}. In the formula (\ref{particlecreation})
$t(\omega_0)$ denotes the transmission amplitude describing the fraction
of the wave that enters the collapsing body, whose collapse produced 
the extreme black hole in \cite{gao}.
 The wavepacket has a spread $\epsilon $ 
in frequencies
around $\omega_0$, and in fact it is the use of such wavepackets that
allows for a consistent calculation of the particle creation
in the extremal black-hole case. The above limit is obtained 
by means of certain analyticity properties of the particle creation 
number~\cite{gao}.

In the case of space-time foam, we have no way (at present) of understanding 
the spontaneous formation of such black holes from the 
quantum gravity vacuum, and hence in our case, one should assume that 
the above results can be extrapolated to this case.
In such a situation, then, $t(\omega_0)$ would be a family of parameters 
describing 
the
space-time foam medium.   

{}From the bounded expression (\ref{particlecreation}),
we observe that since $2n\pi$ represents time, the rate of 
particle creation would drop to zero faster than any (positive) power 
of time at late times. This is in agreement with the abovementioned
considerations about extremal black holes, in particular 
with the absence of particle creation in such a case. 
{}From the smooth connection of non extremal black holes to the 
extremal ones, encountered in string theory~\cite{lifschytz},
we can also conclude that near extremal black holes would
be characterised by relatively small particle creation rate,
as compared with their neutral counterparts.

If we can extrapolate the above-described semi-classical results 
to the quantum gravity foamy ground state, it becomes clear that
microscopic black holes which are near extremal would evaporate 
significantly less, compared with their neutral counterparts. 
Thus, we may assume, that near extremal black holes in the foam
would ``live '' longer, and as a result they would have more time to 
interact with ordinary matter, such as neutrinos. Such charged black
holes would therefore constitute the dominant source of 
charge fluctuations in the foam that could be responsible for 
foam-induced neutrino mass differences according to the idea proposed 
in \cite{bm2}. 

Indeed, the emitted electrons from such black holes,
which as stated above are emitted preferentially, compared
to muons or other charged particles, as 
they are the lightest, 
would then have more time
to interact (via coherent standard model interactions) with the 
electron-neutrino currents, as opposed to muon neutrinos.
This would create a {\it flavour bias} of the foam medium, 
which could then be viewed as the ``quantum-gravitational analogue'' of the 
MSW effect in ordinary media (where, again, one has only electrons, 
since the muons would decay quickly).
In this sense, the quantum gravity medium would be responsible 
for generating effective neutrino mass differences~\cite{bm2}.
Since the charged-black holes lead to a stochastically fluctuating medium,
we shall consider the formalism of the MSW effect for 
stochastically fluctuating media~\cite{loreti}, 
where the density of electrons would be replaced the density of charged black 
hole/anti black hole pairs.

The non-perturbative nature
of quantum gravity foam, makes the above semi-classical
computation unreliable. Hence it 
may not be true 
in a complete theory of quantum gravity.
However, as we shall argue later in this paper, ,
one can already place stringent bounds on the
portion of the neutrino mass differences that 
may be due to quantum gravity foam, as a result of 
current neutrino data.

After this theoretical discussion 
we now proceed to give a brief description of the most important 
phenomenological consequences of such a scenario involving decoherence.
These can help in imposing stringent constraints on the 
percentage 
of the neutrino mass difference that could be due to the quantum-gravity 
medium. For simplicity we restrict ourselves to two generations,
which suffices for a demonstration of the important generic properties
of decoherence.
The extension to three generations is straightforward, albeit 
mathematically more complex~\cite{bmsw}.
The stochasticity of the space-time foam medium 
is best described~\cite{bmsw} by 
including in the time evolution of the neutrino
density matrix a a time-reversal (CPT) breaking
decoherence matrix of a double commutator form~\cite{loreti,bmsw},
\begin{eqnarray}
&&\partial_t \langle \rho\rangle =  L[\rho]~, \nonumber \\
&& L[\rho]=
-i[H + H'_{I},\langle \rho\rangle]-\Omega^2[H'_I,[H'_I,\langle \rho\rangle]]
\label{double} 
\end{eqnarray} 
where $\langle n(r) n(r') \rangle = \Omega^2n_0^2
\delta (r - r') $ denote the stochastic (Gaussian) fluctuations of
the density of the medium, 
and 
 \ba 
H'_I=\left(%
\begin{array}{cc}
  (a_{\nu_{e}}-a_{\nu_{\mu}})\cos^2(\theta)  & (a_{\nu_{e}}-a_{\nu_{\mu}})\frac{\sin2\theta}{2}   \\
  (a_{\nu_{e}}-a_{\nu_{\mu}})\frac{\sin2\theta}{2}  & (a_{\nu_{e}}-a_{\nu_{\mu}})\sin^2(\theta)  \\
\end{array}%
\right) \ea
is the MSW-like interaction in the mass eigenstate basis, where 
$\theta$ is the mixing angle. 
This double-commutator decoherence is a
specific case of Lindblad evolution~\cite{lindblad},
which guarantees complete positivity of the time evolved 
density matrix~\cite{benatti}.

We note at this stage that,
for gravitationally-induced MSW effects (due to, say, black-hole foam
models as in \cite{bm2}), one may denote the difference, between neutrino flavours, 
of the effective interaction strengths, $a_i$,
 with the environment by: 
\ba
\Delta a_{e\mu} \equiv a_{\nu_e}-a_{\nu_\mu} \propto G_N n_0
\ea
with $G_N=1/M_P^2$, $M_P \sim 10^{19}~{\rm GeV}$, the four-dimensional
Planck scale, and 
in the case of 
the gravitational MSW-like effect~\cite{bm2} $n_0$ 
represents the 
density  of charge black hole/anti-black hole pairs.
This gravitational coupling replaces the weak interaction
Fermi coupling constant $G_F$ in the conventional MSW effect.
This is the case we shall be interested in this work.

In such a  case the density fluctuations $\Omega^2$ are therefore assumed
small compared to other quantities present in the formulae, and an expansion to leading order in $\Omega^2$ is in place.
Following then a standard
analysis~\cite{benatti,loreti,bmsw} one obtains the following expression for
the neutrino transition probability $\nu_e \leftrightarrow
\nu_\mu$ in this case, to leading order in the small
parameter $\Omega^2 \ll 1$:
{\small 
\begin{eqnarray}
&&    P_{\nu_e\to \nu_{\mu}}=   \nonumber \\
    && \frac{1}{2} + e^{-\Delta a_{e\mu}^2\Omega^2t(1+\frac{\Delta_{12}^2}{4\Gamma}
(\cos(4\theta)-1))}
    \sin(t\sqrt{\Gamma})\sin^2(2\theta)\Delta
a_{e\mu}^2\Omega^2\Delta_{12}^2
    \left(\frac{3\sin^2(2\theta)\Delta_{12}^2}{4\Gamma^{5/2}}
-\frac{1}{\Gamma^{3/2}}\right) \nonumber \\
    && -e^{-\Delta
    a_{e\mu}^2\Omega^2t(1+\frac{\Delta_{12}^2}{4\Gamma}(\cos(4\theta)-1))}
    \cos(t\sqrt{\Gamma})
\sin^2(2\theta)\frac{\Delta_{12}^2}{2\Gamma}  \nonumber \\
    &&-e^{-\frac{\Delta a_{e\mu}^2\Omega^2 t \Delta_{12}^2\sin^2(2\theta)}{\Gamma}}
    \frac{(\Delta a_{e\mu}+\cos(2\theta)\Delta_{12})^2}{2\Gamma}
\label{2genprob}
\end{eqnarray}}
where $\Gamma= (\Delta a_{e\mu}\cos(2\theta)+\Delta_{12})^2+\Delta
a_{e\mu}^2\sin^2(2\theta)~,$ $\Delta_{12}=\frac{\Delta m_{12}^2}{2p}~.$

{}From (\ref{2genprob}) we easily conclude 
that the exponents of the
damping factors due to the stochastic-medium-induced decoherence,
are of the generic form, for $t = L$, with $L$ the oscillation
length (in units of $c=1$): 
\ba
{\rm exponent} \sim 
-\Delta a_{e\mu}^2\Omega^2 t f(\theta)~;~
f(\theta) = 
1+\frac{\Delta_{12}^2}{4\Gamma}(\cos(4\theta)-1)~, ~{\rm or} ~
\frac{\Delta_{12}^2\sin^2(2\theta)}{\Gamma}
\label{gammadelta} 
\ea 
that is proportional to the stochastic fluctuations of the 
density of the medium.
The reader should note at this stage that, in
the limit $\Delta_{12}\to 0$, which could characterise the situation
in \cite{bm2}, where the space-time foam effects on the
induced neutrino mass difference are the dominant ones, the damping
factor is of the form $ {\rm exponent}_{{\rm gravitational~MSW}}
\sim -\Omega^2 (\Delta a_{e\mu})^2 L~,$ with the precise value of the
mixing angle $\theta$ not affecting the leading order of the various
exponents. However, in that case, as follows from (\ref{2genprob}),
the overall oscillation probability is suppressed by factors
proportional to $\Delta_{12}^2 $, and, hence, the stochastic
gravitational MSW effect~\cite{bm2}, although in principle
capable of inducing mass differences for neutrinos, however does not
suffice to produce the bulk of the oscillation probability, which is
thus attributed to conventional flavour physics.

There are other models of stochastic space-time foam
also inducing decoherence, for instance the 
ones discussed in~\cite{ms,bmsw}, in which one averages
over random (Gaussian) fluctuations of the background space-time metric
over which the neutrino propagates.
In such an approach, one considers merely the Hamiltonian of the neutrino
in a stochastic metric background. The stochastic fluctuations
of the metric would then pertain to the Hamiltonian (commutator) part 
of the density-matrix evolution. In parallel, of course, one should also
consider environmental decoherence-interactions of Lindblad 
(or other) type,
which would co-exist with the decoherence effects due to the 
stochastic metric fluctuations in the Hamiltonian.
For definiteness in what follows we restrict ourselves only 
to the Hamiltonian part, with the aim of demonstrating clearly
the pertinent effect and study their difference from Lindblad decoherence.

In this case, one obtains transition probabilities
with exponential damping factors in front of the oscillatory
terms, but now the scaling with the oscillation length (time) is
quadratic~\cite{ms,bmsw}, consistent with time reversal invariance of the
neutrino Hamiltonian.  For instance, for the two generation case, which
suffices for our qualitative purposes in this work, 
we may consider 
stochastically fluctuating space-times 
with metrics fluctuating along the direction of motion (for simplicity)~\cite{ms}
 \begin{eqnarray} 
    g^{\mu\nu}=\left(
\begin{array}{cc}
  -(a_1+1)^2 + a_2^2 & -a_3(a_1+1) +a_2(a_4+1) \\
  -a_3(a_1+1) +a_2(a_4+1) & -a_3^2+(a_4+1)^2 \\
\end{array}
\right). 
\end{eqnarray} 
with random variables $\langle
a_i\rangle =0$ and $\langle a_i a_j\rangle =
\delta_{ij}\sigma_i$.

Two generation Dirac neutrinos, then, which are considered
for definiteness in \cite{ms} (one would obtain similar results,
as far as decoherence effects are concerned in the Majorana case),
with an MSW interaction 
$V$ (of unspecified origin, which thus 
could be a space-time foam effect) yield the following 
oscillation 
probability:
\begin{eqnarray}
&&\langle e^{i(\omega_1-\omega_2)t}\rangle= e^{i\frac{{\left( {z_0^ +   - z_0^ -  } \right)t}}{k}}
e^{-\frac{1}{2}\left(-i\sigma_1 t\left(\frac{(m_1^2-m_2^2)}{k}+
V\cos2\theta\right)\right)} \times \nonumber \\
&&  e^{-\frac{1}{2}\left(\frac{i\sigma_2t}{2}\left( \frac{(m_1^2-m_2^2)}{k}+V\cos2\theta \right)
    -\frac{i\sigma_3t}{2}V\cos2\theta\right)} \times
    \nonumber \\&& e^{-\left(\frac{(m_1^2-m^2_2)^2}{2k^2}
   (9\sigma_1+\sigma_2+\sigma_3+\sigma_4)+\frac{2V\cos2\theta(m_1^2-m_2^2)}{k}
 (12\sigma_1+2\sigma_2-2\sigma_3)\right)t^2} \label{gravstoch}
 \end{eqnarray}
where $k$ is the neutrino energy, $\sigma_i~, i=1, \dots 4$
parametrise appropriately the stochastic fluctuations of the metric
in the model of \cite{ms}, $\Upsilon  = \frac{{Vk}}{{m_1^2  -
m_2^2 }}$, $\left| \Upsilon  \right| \ll 1$, and $k^2  \gg m_1^2~,~m_2^2 $, and
 \begin{eqnarray}
 z_0^ +   &=& \frac{1}{2}\left(m_1^2  + \Upsilon (1 + \cos 2\theta )(m_1^2  - m_2^2 ) + \Upsilon ^2 (m_1^2  - m_2^2 )\sin ^2 2\theta \right) \nonumber \\
 z_0^ -   &=& \frac{1}{2}\left(m_2^2  + \Upsilon (1 - \cos 2\theta )(m_1^2  - m_2^2 ) - \Upsilon ^2 (m_1^2  - m_2^2 )\sin ^2 2\theta \right)~.
 \end{eqnarray}
Note that the metric fluctuations-$\sigma_i$ induced modifications
of the oscillation period, as well as exponential $e^{-(...)t^2}$
time-reversal invariant damping factors~\cite{ms},
in contrast to the Lindblad decoherence, in which the 
damping was of the form $e^{-(...)t}$. This  feature is
attributed to the fact that in this approach, only the Hamiltonian
terms are taken into account (in a stochastically fluctuating metric
background), and as such time reversal invariance $t \to -t$ is not
broken explicitly. But there is of course decoherence, and the
associated damping.

A few remarks are now in order regarding the 
similarity of this latter type of decoherence (\ref{gravstoch})
with the one mimicked~\cite{ohlsson} by ordinary 
uncertainties in neutrino experiments 
over the precise energy $E$ of the beam (and in some cases over the
oscillation length $L$).
Indeed, consider the Gaussian average of a generic 
neutrino oscillation probability  
over the L/E
dependence  
 $\langle P \rangle = \int_{-\infty}^{\infty} dx P(x) \frac{1}{\sigma
    \sqrt{2\pi}}e^{-\frac{(x-l)^2}{2\sigma^2}}~,$
with $l=\langle x \rangle$  and $\sigma=\sqrt{\langle (x-\langle
x \rangle )^2\rangle }$, $x=\frac{L}{4E}$, and assuming
the independence of  $L$ and $E$, which allows to write $l=\langle L/E\rangle
=\langle L\rangle /4\langle E\rangle$.
A pessimistic and an optimistic upper bound for
$\sigma$ are given by~\cite{ohlsson}
\begin{itemize}
    \item pessimistic:  $\sigma \simeq \Delta x =\Delta \frac{L}{4E} \leq
    \Delta L \big{|}\frac{\partial x}{\partial L}\big{|}_{L=\langle L\rangle,
    E=\langle E \rangle}+\Delta E \big{|}\frac{\partial x}
    {\partial E}\big{|}_{L=\langle L\rangle,
    E=\langle E \rangle}$

    $=\frac{\langle L\rangle}{4\langle E\rangle}
    \left( \frac{\Delta L}{\langle L\rangle}+\frac{\Delta E}
    {\langle E\rangle}\right)$
    \item optimistic: $\sigma \le \frac{\langle L\rangle}{4\langle E\rangle}
    \sqrt{\left( \frac{\Delta L}{\langle L\rangle}\right)^2+\left(\frac{\Delta E}
    {\langle E\rangle}\right)^2} $
\end{itemize}
Then, 
it is easy to arrive at the 
expression~\cite{ohlsson}
{\small 
\begin{eqnarray} 
&& \langle P_{\alpha\beta} \rangle = \delta_{\alpha\beta} 
- \nonumber \\
&& 2 \sum_{a=1}^n\sum_{\beta=1, a<b}^n{\rm Re}\left(U_{\alpha a}^*
U_{\beta a}U_{\alpha b}U_{\beta b}^*\right)
\left( 1 - {\rm cos}(2\ell \Delta m_{ab}^2)
e^{-2\sigma^2(\Delta m_{ab}^2)^2}\right)
\nonumber \\
&& -2 \sum_{a=1}^n \sum_{b=1, a<b}^n {\rm Im}\left(U_{\alpha a}^*
U_{\beta a}U_{\alpha b}U_{\beta b}^*\right)
{\rm sin}(2\ell \Delta m_{ab}^2)
e^{-2\sigma^2(\Delta m_{ab}^2)^2}
\label{uncert}
\end{eqnarray}}
with $U$ the appropriate mixing matrix.
Notice the $\sigma^2$ damping factor of neutrino oscillation
probabilities, which has the similar form  in terms of the 
oscillation length dependence ($L^2$ dependence) 
as the corresponding damping factors due to the stochasticity 
of the space-time background in (\ref{gravstoch}).  
It is noted, however, that here 
$l$ has to do with the sensitivity of the experiment, and thus the physics
is entirely different.

In the case of space-time stochastic backgrounds, one could still
have induced uncertainties in $E$ and $L$, which however are 
of fundamental origin, and are expected to be more suppressed
than the uncertainties due to ordinary physics, described above. 
Apart from
their magnitude, 
their main difference from the uncertainties in (\ref{uncert}) 
has to do with the specific dependence of the corresponding $\sigma^2$ 
in that case on both $E$ and $L$. 
For generic space-time foam models it is expected that 
an uncertainty in $E$ or $L$ due to the ``fuzziness'' of space time
at a fundamental (Planckian) level will increase with the energy 
of the probe, $\delta E/E, \delta L/L \propto (E/M_P)^\alpha $, 
$\alpha > 0$, 
since the higher the energy the bigger the disturbance (and hence back reaction)on the space time medium. In contrast, ordinary matter effects 
decrease with the energy of the probe~\cite{ohlsson,bow}. 

\section{Fitting the data and attempts to interpret them}

In \cite{bmsw} a three generation Lindblad decoherence model of neutrinos
has been compared against all available experimental data,
taking into account the recent results from KamLand experiment~\cite{kamland}
indicating spectral distortions. 

\begin{figure}
\centering 
\epsfig{figure=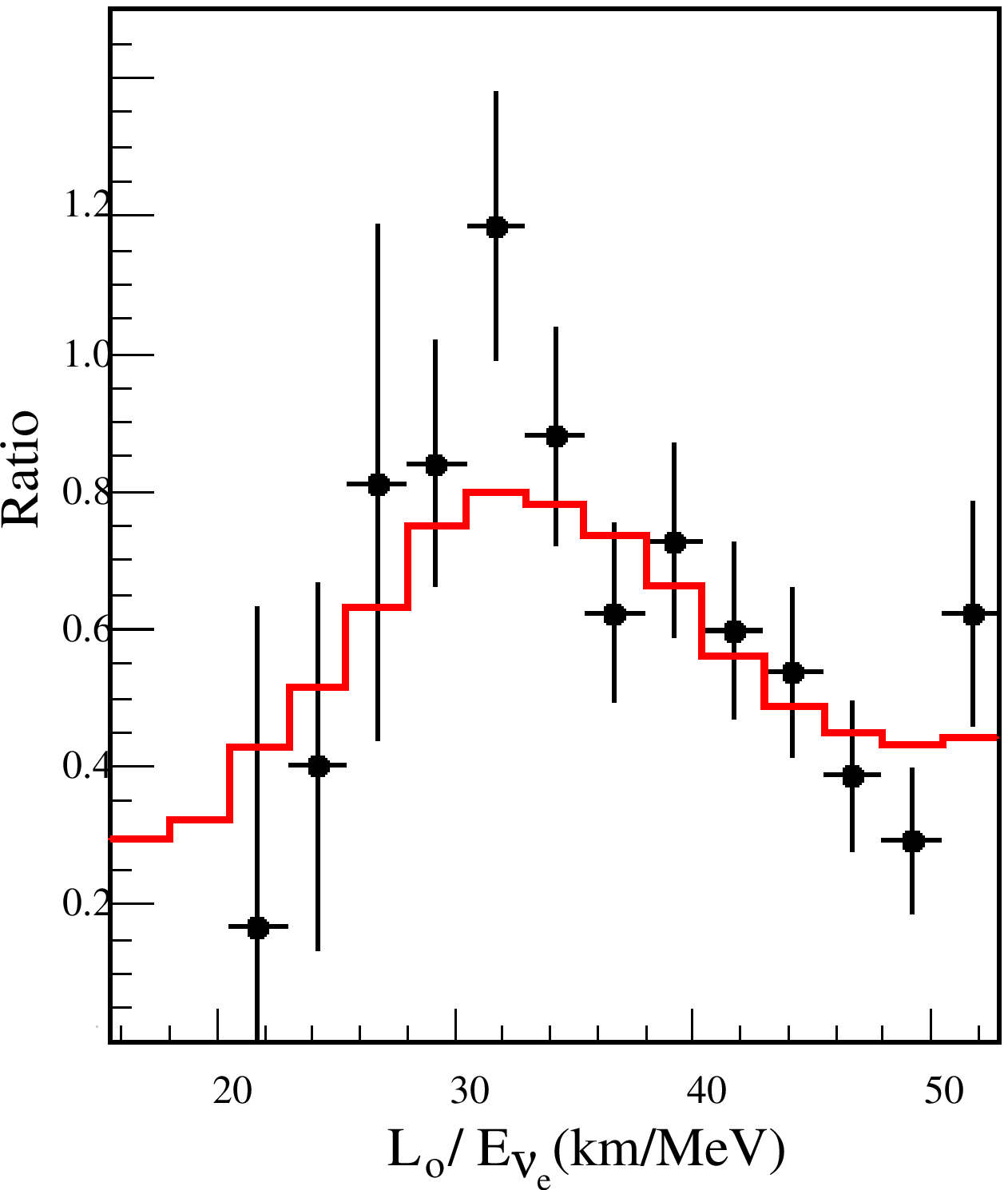,width=7.0cm}
\epsfig{figure=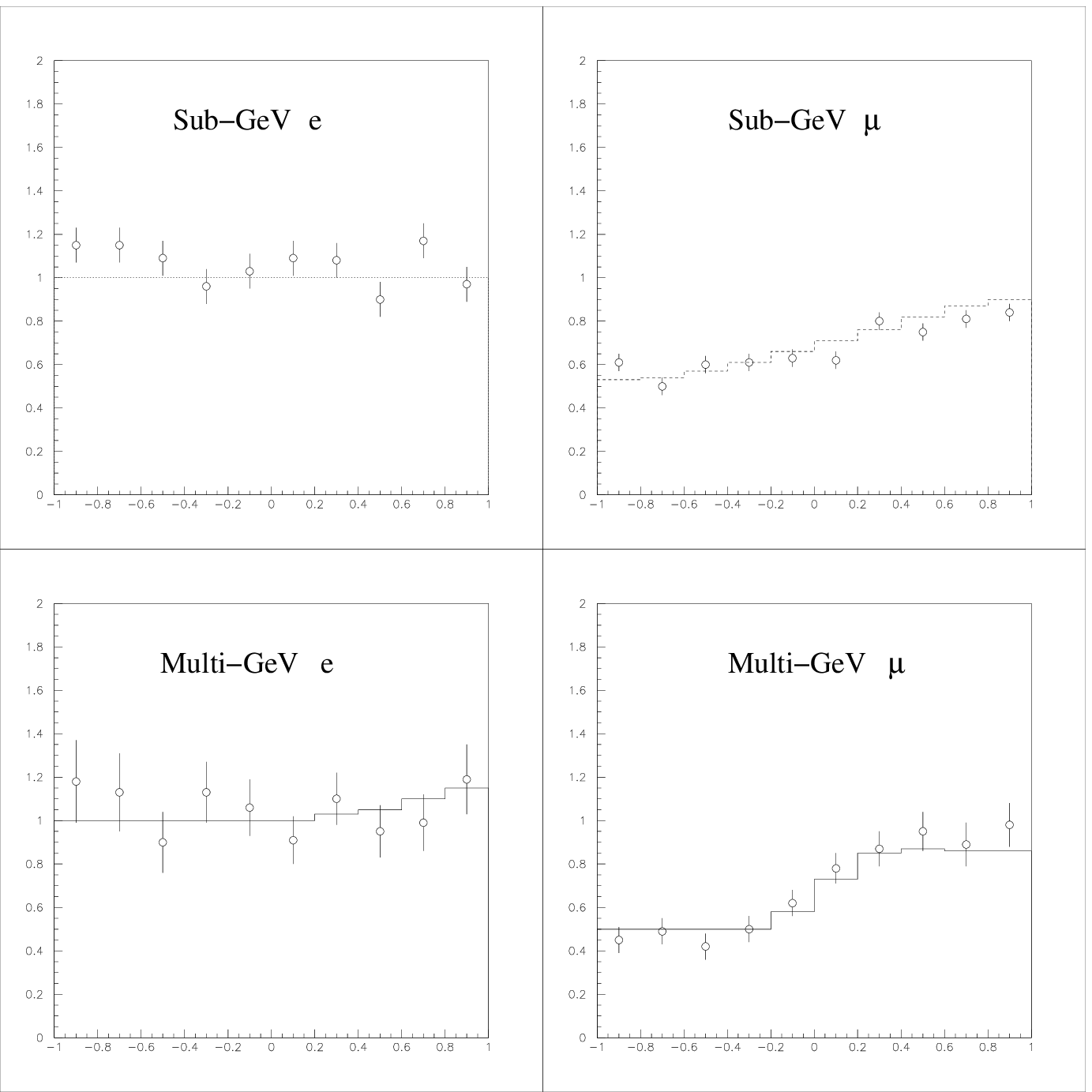,width=8.0cm}
\caption{\underline{Left}: Ratio of the observed $\overline{\nu}_e$ spectrum to the expectation
versus $L_0/E$ for our decoherence model. The dots correspond to KamLAND data.
\underline{Right}:Decoherence fit. The dots correspond to SK data.}
\label{fig1}
\end{figure}

The results are summarised in Fig.~\ref{fig1}, 
which demonstrates the agreement (left) of our model 
with the KamLand spectral 
distortion data~\cite{kamland}, and our best fit (right) 
for the Lindblad decoherence model used in ref.~\cite{bmsw}, 
and in  
Table 1, where we
present the $\chi^2$ comparison for the model in question
and the standard scenario. 

\begin{table}[h]
\centering
\begin{tabular}{|c|c|c|}
\hline\hline
$\chi^2$ &  decoherence  & standard scenario  \\ [0.5 ex]
\hline\hline
SK  sub-GeV& 38.0  & 38.2 \\ \hline
SK Multi-GeV & 11.7  & 11.2 \\ \hline
Chooz & 4.5 & 4.5  \\\hline
 KamLAND & 16.7 & 16.6 \\\hline
LSND & 0.  & 6.8  \\\hline
TOTAL & 70.9  & 77.3 \\[1ex]
\hline\hline
\end{tabular}
\caption{$\chi^2$ obtained for our model and the one obtained in the standard
scenario for the different experiments calculated with the same program.}
\end{table}

The best fit has the feature that only some of the 
oscillation terms in the three generation 
probability formula have non trivial damping factors, 
with their 
exponents being {\it independent} of the oscillation 
length, 
specifically~\cite{bmsw}. If we denote those non trivial 
exponents as ${\cal D}\cdot L$, we 
obtain from the best fit of \cite{bmsw}:
\ba
{\cal D}=- \frac{\;\;\; 1.3 \cdot 10^{-2}\;\;\;}
{L},
\label{special}
\ea
in units of 1/km with $L=t$ the oscillation length. The
$1/L$-behaviour of ${\cal D}_{11} $, implies, as we mentioned, 
oscillation-length independent Lindblad exponents. 

In \cite{bmsw} an analysis of the two types of the theoretical models
of space-time foam, discussed in section 2, has been performed 
in the light of the result of the fit (\ref{special}).   
The conclusion was that the model of the 
stochastically fluctuating media 
(\ref{2genprob}) 
(extended appropriately to three generations~\cite{bmsw}, so as 
to be used for comparison with the real data) 
cannot provide the full explanation for the fit, for the following reason: 
if the decoherent result of the fit (\ref{special}) was exclusively 
due to this model, then the pertinent 
decoherent coefficient in that case, for, say, the
KamLand  experiment with an $L \sim 180$~Km,       
would be $ |{\cal D}| = \Omega^2 G_N^2 n_0^2 \sim 2.84 \cdot
10^{-21}~{\rm GeV}$ (note that the mixing angle part does not affect the
order of the exponent). Smaller values are found for longer $L$,
such as in atmospheric neutrino experiments. 
The independence of the
relevant damping exponent from the oscillation length, then, as required
by (\ref{special}) may be understood as follows in this context: 
In the spirit of \cite{bm2}, 
the quantity $G_N n_0 = \xi \frac{\Delta m^2}{E}$,
where $\xi \ll 1$ parametrises the contributions of the foam to the
induced neutrino mass differences, according to our discussion
above. Hence, the damping exponent becomes in this case $ \xi^2
\Omega^2 (\Delta m^2)^2 \cdot L /E^2 $. Thus, for oscillation
lengths $L$ we have 
$L^{-1} \sim \Delta m^2/E$, and one is left with  the following 
estimate for the dimensionless quantity $\xi^2
\Delta m^2 \Omega^2/E \sim 1.3 \cdot 10^{-2}$. This  
implies that the quantity $\Omega^2$ is proportional to the
probe energy $E$. In principle, 
this is not an unreasonable result, and it is in
the spirit of \cite{bm2}, since back reaction effects onto
space time, which affect the stochastic fluctuations $\Omega^2$, are
expected to increase with the probe energy $E$. However, 
due to the smallness of the quantity $\Delta m^2/E$, for energies 
of the order of GeV, and $\Delta m^2 \sim 10^{-3}$ eV$^2$, 
we conclude (taking into account that 
$\xi \ll 1$) that $\Omega^2$ in this case 
would be unrealistically large
for a quantum-gravity effect in the model. 

We remark at this point that, in such a model,
we can in principle bound independently the $\Omega$
and $n_0$ parameters by looking at the modifications induced by the
medium in the arguments of the oscillatory functions of the
probability (\ref{2genprob}), that is the period of oscillation.
Unfortunately this is too small to be detected in the above example,
for which $\Delta a_{e\mu} \ll \Delta_{12}$.

The second model (\ref{gravstoch}) of stochastic space time
can also be confronted with the data, since in that case
(\ref{special}) 
would imply for the pertinent damping exponent

\ba
&& \left(\frac{(m_1^2-m^2_2)^2}{2k^2}
   (9\sigma_1+\sigma_2+\sigma_3+\sigma_4)+
\frac{2V\cos2\theta(m_1^2-m_2^2)}{k}
 (12\sigma_1+2\sigma_2-2\sigma_3)
\right)t^2 \nonumber \\
&& \sim 1.3  \cdot 10^{-2}~.
\ea 
Ignoring subleading MSW effects $V$, for simplicity,
and considering oscillation lengths $t=L \sim
\frac{2k}{(m_1^2-m^2_2)}$, we then observe that the independence of
the length $L$ result of the experimental fit, found above, may be
interpreted, in this case, as bounding the stochastic fluctuations
of the metric to $9\sigma_1+\sigma_2+\sigma_3+\sigma_4 \sim
1.3. \cdot 10^{-2}$. Again, this is too large to be a quantum gravity
effect, which means that the $L^2$ contributions to the
damping due to stochastic fluctuations of the metric,
as in the model of \cite{ms} above (\ref{gravstoch}), 
cannot be the 
exclusive explanation of the fit.

The analysis of \cite{bmsw} also demonstrated that, at least as 
far as an order of magnitude of the effect is concerned,
a reasonable explanation of the order of the damping 
exponent (\ref{special}), is provided by 
Gaussian-type energy fluctuations, due to 
ordinary physics effects, leading to decoherence-like damping 
of oscillation probabilities of the form 
(\ref{uncert}). The order of these fluctuations, 
consistent with   
the independence of the damping exponent
on $L$ (irrespective of the power of $L$), 
is
\ba
 \frac{\Delta E}{E} \sim 1.6 \cdot 10^{-1}
\ea
if one assumes that this is the principal reason for the
result of the fit.

However, not even this can be the end of the story, given that the 
result (\ref{special}) pertains only to {\it some} of the 
oscillation terms and not all of them, which would be the case expected
for the ordinary physics uncertainties (\ref{uncert}). 
The fact that the best fit 
model includes terms which are not suppressed at all calls for 
a more radical explanation of the fit result, and the issue is 
still wide open. 
It is interesting, however, that the current neutrino data can 
already impose stringent constraints on quantum gravity models, and exclude
some of them from being the 
exclusive source of decoherence, as we have discussed above. 

Coming back to our main point of discussion in this work,
we stress once more that, 
within the classes of stochastic models we discussed in this work, 
one can safely exclude the possibility that 
space-time foam can be at most responsible only for a small part 
of the observed neutrino mass difference, 
and certainly the foam-induced decoherence cannot be the sole reason 
for the result of the best fit (\ref{special}), 
pertaining to a global analysis of the currently available neutrino 
data. Of course, this is not a definite conclusion because one cannot 
exclude the possibility of other classes of theoretical models
of quantum gravity, which could escape 
these constraints. At present, however, we are not aware of any such 
theory.

\section{Neutrino Mass differences, Mixing, Space-time Foam 
and the Cosmological Constant}

Since quantum-gravity decoherence can still be
accommodated by 
the current 
data, despite the above conclusions on the smallness of the 
percentage of the observed neutrino mass difference due to 
the space-time foam medium, we would like in this section to 
speculate on possible implications of these effects to the dark energy 
budget of our Universe~\cite{bm2,mavrodice}.  

In this respect with mention that an approach was suggested 
in \cite{vitiello} for applying a Fock space quantisation 
to field theories with mixing. Their formalism, which was performed
in {\it flat} space time field theories, involved the definition of a 
new type of Fock-space vacuum, the ``flavour vacuum'', $|0(t)\rangle_f$.
This vacuum 
was not connected with the mass eigenstate vacuum, $|0(t)\rangle_m$,
by a unitary transformation in the field theoretic (thermodynamic) limit,
where the volume of the system was taken to infinity. 
Instead, there is a non-unitary transformation $G$, connecting 
these vacuum states, which reads~\cite{vitiello}  
\begin{equation}
|0(t)\rangle_f = G^{-1}_\theta (t)|0(t)\rangle_m~,~~
G_\theta (t) = {\rm exp}\left(\theta \int d^3x [\nu_1^\dagger (x)
\nu_2 (x) - \nu_2^\dagger (x)
\nu_1(x)]\right)~,
\label{nonunit}
\end{equation}
where $\theta$ is the mixing angle, $t$ is the time, and the suffix f(m)
denotes flavour(energy) eigenstates, 
A Bogolubov transformation was necessary to connect 
the creation and annihilation operator 
coefficients appearing in the energy eigenstates 
with the corresponding ones for  
flavour eigenstates, 
which leads naturally to particle creation, and a 
``flavour condensate'' 
\begin{equation}
V_{\vec k} = 
|V_{\vec k}|e^{i(\omega_{k,1} +\omega_{k,2})t}~,
\end{equation} 
with $\omega_{k,i}=\sqrt{k^2 + m_i^2}~,$
$_f\langle 0 |\alpha_{{\vec k}, i}^{r \dagger}\alpha_{{\vec k}, i}^{r}
|0\rangle_f = _f\langle 0 |\beta_{{\vec k}, i}^{r \dagger}
\beta_{{\vec k}, i}^{r}|0\rangle_f = {\rm sin}^2\theta |V_{\vec k}|^2$ 
in the two-generation scenario where we 
restrict our discussion for brevity (in the 
three-generation case there are various $V_{ij}$). 

The Flavour Condensate 
$|V_{\vec k}| =0$ for $m_1=m_2$, and it has a maximum 
at $k^2 =m_1m_2$, and for $k \gg \sqrt{m_1m_2}$
exhibits the following asymptotic behaviour~\cite{vitiello}
\begin{equation}
|V_{\vec k}| \sim \frac{(m_1 - m_2)^2}{4|{\vec k}|^2}, \quad 
k \equiv |{\vec k}| \gg \sqrt{m_1m_2}
\end{equation}

The analysis of \cite{henning} went one step further to claim 
that the mass eigenstate vacuum was not the appropriate one to 
conserve probability, and hence the only appropriate  
vacuum was the flavour one, 
which respected this property, but which involved a 
modified expression for the probability, containing 
terms proportional to the  flavour condensate.

Using this vacuum as the physical one has important 
cosmological consequences. Indeed, 
computing the flavour-vacuum average of 
energy-momentum tensor $T_{\mu\nu}$ 
of a Dirac (for definiteness, although the Majorana case also leads 
to similar results) fermion field in a Robertson-Walker
background space-time, leads for the temporal 
component $T_{00}$~\cite{vitiello}: 
\begin{eqnarray}
&& _f\langle 0|T_{00} |0\rangle_f 
=\langle \rho_{\rm vac}^{\rm \nu-mix}\rangle \eta_{00} \equiv \Lambda \eta_{00}
\nonumber \\
&& = \sum_{i,r}\int d^3 k \omega_{k,i}\left(_f\langle 0 |\alpha_{{\vec k}, i}^{r \dagger}\alpha_{{\vec k}, i}^{r}|0\rangle_f + 
_f\langle 0 |\beta_{{\vec k}, i}^{r \dagger}\beta_{{\vec k}, i}^{r}|0\rangle_f\right) 
\nonumber \\
&& = 8{\rm sin}^2\theta \int_{0}^{K} d^3k (\omega_{k,1} +  
\omega_{k,2})|V_{\vec k}|^2.
\end{eqnarray}
where $\eta_{00}=1$ in a Robertson-Walker (cosmological) metric background.

A consistent, and physically relevant choice of the cutoff scale
has been proposed in \cite{bm2} to be 
 $K \equiv k_0 \sim m_1 + m_2$.
This choice
is compatible with a decoherence-induced mass difference scenario, 
since it implies that only the infrared 
neutrino modes, with momenta less than the typical mass scales
$m_1 + m_2$, feel mostly 
the space-time medium effects, since, being slow, 
they have more time to interact with 
the gravitational environment.

For hierarchical neutrino models with $m_1 \gg m_2$ $ \to$  
$k_0 \gg \sqrt{m_1m_2}$,
modes near the cutoff give the dominant 
contributions to the vacuum energy $\Lambda$ 
(due to the divergences involved), 
\begin{eqnarray} 
&&\Lambda \equiv \langle \rho_{\rm vac}^{\rm \nu-mix}\rangle 
\sim 
8\pi{\rm sin}^2\theta (m_1 - m_2)^2 (m_1 + m_2)^2\times \nonumber \\ 
&&\left(\sqrt{2} + 1 +{\cal O}(\frac{m_2^2}{m_1^2})\right) \propto 
{\rm sin^2}\theta (\Delta m^2)^2
\label{desitternu}
\end{eqnarray}
This implies that the mixing and mass difference for neutrinos
lead to a contribution to the cosmological constant (or better 
dark energy 
of the vacuum) of the phenomenologically right order of magnitude.

There are several issues with the above scenario that need to be 
addressed, before the above considerations are accepted.
The first and most important of all is the fact that these calculations
have been performed in a flat space time, but the result 
(\ref{desitternu}) implies a curved de Sitter space time.
Moreover, in \cite{vitiello} the mass difference of neutrinos 
was assumed from the beginning, although in the approach of \cite{bm2,ms,bmsw},
there is a dynamical component which is due to 
the (non flat at microscopic scales) space-time foam vacuum.

In view of the particle creation characterising the flavour vacuum 
in the approach of \cite{vitiello}, our stochastic space time 
model may be the most appropriate framework where 
such issues can be discussed in a mathematically and physically
consistent setting. In addition to the modifications 
to the oscillation probability appearing in the approach 
of \cite{henning,vitiello}, in our case there are the CPT violating 
decoherence
modifications (damping), which imply a microscopic time irreversibility, 
and a non-unitary evolution. Moreover, 
as discussed in \cite{masa2}, the presence 
of dark energy contributions in space time imply additional
damping factors in the oscillation probability. 
In this respect, the non unitarity
involved in the definition of the flavour vacuum (\ref{nonunit}) 
may acquire important physical meaning.

Several other issues deserve careful study, among which 
a detailed and proper study, using curved space time techniques,
of the equation of state characterising the neutrino fluid
in the presence of the foam.
The flat space time attempt of \cite{vitiello2} 
is not, in our opinion, sufficient to give a complete and consistent
answer for this specific question, which is important to 
cosmologists.  
We hope to come back to such important questions in the near
future.

\section{Conclusions and Outlook}

This conference celebrated 50 years from the neutrino 
discovery. Ever since its discovery this elusive particle 
keeps surprising us. At first, scientists thought that 
energy was not conserved in the nuclear $\beta$-decays, 
before Pauli makes the decisive suggestion on the 
presence of the neutrino. 

In the standard model of particle physics the neutrino 
appears massless, but during the last decade a plethora
of delicate experiments have shown unambiguously 
(albeit indirectly) the existence of a neutrino mass, by measuring 
oscillations, thereby allowing for estimates
of the mass differences. 

The origin of such a mass and mass differences are still 
major issues, and intense research is at present under way
in order to tackle such questions.
In refs.~\cite{bm2,bmsw,ms}, 
we have put forward a conjecture, 
reviewed in this talk, 
according 
to which  
part of the observed mass differences of neutrinos might be due to
completely new physics, that of Quantum Gravity.
As we have discussed in this talk, 
stringent constraints can be imposed by the current 
data on the proportion of the neutrino mass difference that could be 
due to such effects. Nevertheless, experiments
are still compatible with the presence of a space time 
foam medium, responsible for neutrino decoherence
and generation of part of the mass differences between neutrino flavours.

This brings in other interesting scenaria on 
possible links of neutrinos
with a dark energy component of the Universe~\cite{mavrodice,vitiello}
of an unconventional origin, consistent with the current 
phenomenology. Interesting questions as to what type of quantum fluid 
neutrinos actually constitute, 
are still probably far from being completely understood,
in view of the possible mixing of neutrinos with  the quantum-gravity foam.
A peculiar proportionality 
relation between a Dark Energy component of the Universe and the 
product of the sum of the neutrino mass differences 
$\Delta m^2$ times some trigonometric factors  
of the mixing angle has been proposed, which stills appears 
compatible with the data. All these issues can be understood rigorously 
only if one formulates the problem of quantum field theory mixing 
in curved (de Sitter) space times. 

In view of these considerations, 
Neutrino Physics may provide a very useful guide
in our quest for a theory of Quantum Gravity, in particular 
stringent constraints on CPT Violation (or better, 
microscopic time irreversibility). 
As discussed above, the latter may not be 
an academic issue, but a real feature of 
Quantum Gravity. As we have reviewed in this work, 
the scenario of 
three-generation neutrino decoherence 
plus mixing~\cite{bmsw}
is still compatible with all the existing neutrino data, 
including KamLAND spectral distortions. It yields decoherence
damping factors of a peculiar behaviour (independent of the 
oscillation length),  
which still calls
for a rigorous explanation. 

Clearly neutrino physics 
hides many more surprises and mysteries, some of which may be revealed
already in the next round of experiments.
Who knows what opinion about these fascinating particles
scientists would have formed by the date 
we shall celebrate the centenary 
of the experimental neutrino discovery.
One thing is certain, though, that much more work, both theoretical
and experimental, should be done before definite conclusions 
are reached concerning the precise nature and properties of
neutrinos.

\section{Acknowledgements}

N.E.M. would like to thank Prof. M.~Baldo-Ceolin for the invitation
to speak in this conference celebrating {\it Fifty Years from the Neutrino 
Experimental Discovery}, and for providing a 
very interesting and thought stimulating 
meeting. We would also like to 
thank G. Barenboim and A. Waldron-Lauda for an enjoyable collaboration,
results of which have been reviewed here, and V. Mitsou 
for informative discussions on experimental data.
The work of N.E.M. is partially supported by funds made available 
by the European Social Fund (75\%) and National (Greek)
Resources (25\%) - EPEAEK B - PYTHAGORAS.

\end{document}